\documentstyle[12pt,aasms4]{article}

\begin{document}

\title{\it{Letter to the Editor}}\hfill

\vskip.5truecm

\title{A MORPHOLOGICAL METHOD TO DETERMINE CO-ROTATION 
       RADII IN SPIRAL GALAXIES}

\author{Iv\^anio Puerari}

\affil {Instituto Nacional de Astrof\'{\i}sica, \'Optica y Electr\'onica,
 Apartado Postal 216, 72000 Puebla, Mexico}

\author{and}

\author{Horacio Dottori}

\affil{Departamento de Astronomia, Instituto de F\'{\i}sica, UFRGS,  
	CxP 15051, CEP 91501-907  Porto Alegre, Brazil}

\begin{abstract} 

Shock induced star formation in a stellar density wave scenario 
produces an azimuthal gradient of ages across the spiral arms
which has opposite signs on either side of the corotation resonance (CR). 
We present a method based on  the Fourier analysis of azimuthal profiles,  
to locate the CR and  determine the arm character
(trailing or leading) in spiral galaxies. Basically, we compare the
behavior of 
the  phase angle of the two-armed spiral in blue and infrared colors 
which pick out respectively young and older disk stellar population. 
We illustrate  the method using theoretical leading and trailing, 
spirals. We have
also applied the method to the spiral galaxies NGC\,7479, for which
we confirm the reported leading arms, and NGC\,1832. In these galaxies 
we find two and three CRs respectively.

\end{abstract}

\keywords{galaxies: spiral -- galaxies: kinematics and dynamics
-- galaxies: individual (NGC\,1832, NGC\,7479) -- method: analytical}

\title{Introduction}

  Schweizer (1976) and Beckman \& Cepa (1990, hereinafter BC90) have previously 
discussed what would be the behavior of the colors across spiral arms
if a shock generated by a spiral density wave (SDW) induces  star formation.
The main azimuthal observable  characteristics
of this scenario are steeper azimuthal profiles and bluer color
indexes on the side where the shock front is located.
 Elmegreen, Elmegreen \&
Montenegro (1992) also pointed out that such evidence for the CR is
clear in gas-rich galaxies in the form of sharp endpoints to
star formation ridges and dust lanes in two arm spirals.    

 BC90 analyzed azimuthal profiles of the
spiral galaxies NGC\,7479 and
NGC\,1832,  detecting a systematic shift between the arm phase in
B and I colors. For the B band, OB stars represent 70\% of the luminosity
and the older stars of the disk only 30\%. For the I band, most of the
contribution (80\%) is due to disk stars (Schweizer 1976). Then, 
within the framework of the SDW theory,  BC90 have  basically 
detected the relative position of the shock front with respect 
to the SDW perturbation.
In a previous paper (Puerari \& Dottori, 1992), we proposed a
method to determine the leading or trailing character of perturbation
in spiral galaxies by analyzing the distribution of HII regions. 
We present in this  {\it Letter} a useful method  to locate CR and to determine
the leading or trailing character of a perturbing pattern. It is based
on the analysis of azimuthal profiles of 
two-armed spiral galaxies by means of Fourier transforms
applied to B and I images.  

The method is described in the next section, where we also
analyze ideal spirals, and show the behavior of the phase $\Theta$
in the $\theta\,{\rm vs}\,r$ diagrams,
suitable to be compared directly with those of true spiral galaxies.
In Sect.\,3 we discuss the case of NGC\,7479, reported as  
a leading spiral (BC90), and that of NGC\,1832. Finally, our
conclusions are given in Sect.\,4.

\title{The method}

Shock induced star formation in a stellar density wave scenario 
produces an azimuthal spread of ages across the spiral arms. At the 
corotation radius (CR) the angular velocity of the perturbation ($\Omega_p$) 
and that of the stellar disk ($\Omega$) coincide. A comoving observer
at the CR will see  out- and inwards,  the shock front to change from
one side of the spiral to the other,  consequently 
reversing  the order in which young and older 
disk stellar populations appear in azimuthal profiles across the arm.
In order to detect the shock front jump, 
we analyze  the relative behavior of the SDW and shock front phases, 
$\Theta_{dw}$ and 
$\Theta_{sf}$ respectively, by means of the Fourier transform of azimuthal profiles  $I_r(\theta)$  given by,

  $$ {\cal{F}}_2(r)=\int_{-\pi}^{\pi} I_r(\theta) e^{-2i\theta} d\theta$$

The phase $\Theta$ can be obtained as,

$$\Theta(r)= {\rm tan}^{-1}\,{{ {Re}[{\cal{F}}_2(r)}]\over{{Im} [{\cal F}_2(r)}]}$$ 

\noindent where $Re$ and $Im$ mean the real and imaginary 
part of the complex Fourier coefficient.
 
In Figs.\,1a to 1d, we show idealized two-armed spirals of 
leading and trailing character, and on-the-sky views of types S and Z. 
The light line represents the SDW and the heavy line the shock front.
Figs.\,2a to 2d show the relative behavior of $\Theta_{sf}(r)$ 
and $\Theta_{dw}(r)$ for the four cases of Fig.\,1.
As can be seen, the position of the CR and the trailing or leading character
of the arms are clearly revealed in these plots.

\title{Analysis of NGC\,1832 and NGC\,7479}
 
Frames are oriented  with N at the top and
E to the  left. The azimuthal profiles are obtained  
counterclockwise, beginning at the South.

The B and I frames of NGC\,1832 images are the same as those used in BC90. 
NGC\,7479 images have
been obtained later on, also with the J. Kapteyn telescope, at
the Roque de los Muchachos Observatory in La Palma, 
with a larger CCD, which allowed  the whole galaxy to be imaged (The
images were kindly provided by J. Cepa). 

Photometrically calibrated frames are not required to carry 
out the phase  analysis, but it was necessary to eliminate
all the stars from the frames and to deproject the galaxies. 

Following the discussion of the two previous sections,  the curves 
$\Theta_B(r)$ and $\Theta_I(r)$ are respectively equivalent to  
$\Theta_{sf}$ and $\Theta_{dw}$.

NGC\,7479 presents an on-the-sky view of Z-type . 
The curves $\Theta_B(r)$ 
and $\Theta_I(r)$ (Fig.\,3), intercept at $r\approx 55"$, which coincides
with the CR position given by BC90.  The relative position of
the curves is comparable to that
of Fig.\,2c, which confirms, consequently,  the leading 
character of the pattern with CR at the end of the bar. A second 
intersection is present  at $r\approx23"$, that time indicating 
a trailing pattern. 
As Fig.\,4 shows, the outer CR  lies on 
the northern extreme of the bar and is about 20\% outwards from  the 
southern one. Fig.\,4 also shows that the inner CR 
bounds the fat internal part of the bar, and
marks a cut in the fine internal dust lane that comes out of
the nucleus. These features are signatures of a CR, as pointed
out by Elmegreen, Elmegreen \& Montenegro (1992). 
The presence of a second CR inside a rigid bar might be due to
the perturbation of the velocity field caused by the 
strong mass inflow, with speed as high as 20\,km/sec
(Quillen et al. 1995). It is important to point out,
as numerical simulations show,
that a inflow speed as high as this is consistent with 
the predictions of galaxy mergers (Mihos \& Hernquist 1994a,b) 
and not with that of a bar perturbation in an 
isolated galaxy (Athanassoula 1992, 1-6 km/sec).
Beckman suggests that there could exist a trailing solution
for NGC\,7479 also. However, Beckman realized this scenario is not easy
to verify. Since we have not found a third CR, like that in NGC\,1832 (see Fig.\,7),
Beckman's claim would require that $\Omega_p(r)\, \leq\, \Omega(r)$ for
$r\,\geq\,R_{CR_2}$. Spatially larger and deeper images would be necessary in order
 to check the existence of a third CR beyond the limits of the
images used in this paper.

NGC\,1832  presents an on-the-sky view of S-type. 
The curves  $\Theta_B(r)$ 
and $\Theta_I(r)$ (Fig.\,5) present  three intersections. 
This plot also shows that the structure inside the ring 
($r\leq 17"$) is not precisely a bar, but an arm-like
structure, with a winding opposite to the external arms. The
first intersection indicate a  CR coincident with the internal border
of the ring ($r\approx 15"$,  Fig.\,6). 
The second and third  CRs, mark gaps in the strong eastern arm 
and the broadening and bifurcation of the western one. 
The  inner and the outer CRs indicate  leading patterns, and 
the intermediate CR a trailing one. In Fig.\,7 we show 
schematically,  the relative behavior of the disk
and the perturbing pattern  angular velocities. 

The errors in the determination of the CRs are difficult to
evaluate in real cases and to estimate by models such as those of
Fig.\,1. Factors that may affect this determination are the departure from
180$^\circ$ symmetry of the two arms structure, differences between the
brightness of the two arms, the pitch angle (bars, and open arms will give more
reliable results that closed ones), the arm width, etc. 
At all events, we suggest
that the method should not be applied to galaxies with inclination larger
than 55$^\circ$ to 60$^\circ$.
 
\title{Conclusions}

We present a method  to locate the CR and  determine the trailing or leading
character of the spiral pattern in  two-armed spiral galaxies, 
based on the Fourier transform of azimuthal profiles
in B and I images.

We present also theoretical plots of the relative 
behavior of the phase
angles $\Theta_B(r)$ and $\Theta_I(r)$ in the $\theta\,{\rm vs}\,r$ plane,
for trailing and leading spirals that  show up as S- or Z-type, useful
to easily and  quickly check  the result when applying this
method to real spiral galaxies.

We confirm for  NGC\,7479 the existence of the leading  pattern 
with CR at the extreme of the bar. We found in this 
galaxy the existence of a internal CR, indicating a trailing pattern.

NGC\,1832 presents three CRs, the inner and the outer ones indicate
leading pattern and the intermediate CR a trailing one. The most plausible 
physical interpretation for this situation is the existence of two
pattern speeds.

The continuity
of the curves $\Theta_B$ and $\Theta_I$  in both galaxies 
might be indicating 
that trailing and leading phenomena are physically related
in these objects. A larger sample of galaxies is being analyzed.

We acknowledge Dr. Beckman for helpful suggestions about the interpretation
of the phase diagrams specially for Figure\,7. This work is being
partially supported by the brazilian institution CNPq and CAPES and CONACYT from
Mexico.

\newpage

\begin{figure}
\caption{The position of the shock front (heavy line) with respect to the
SDW (light line) for a: a- leading, S-type; b- trailing, S-type;
c-leading, Z-type; d- trailing, Z-type wave. The arrow in each panel
indicates the sense of the disc rotation. 
Note: An S-type object would be seen as Z-type if viewed from the 
opposite pole of its axis of rotation. The distinction is 
observational rather than physical.}

\end{figure}

\begin{figure}
\caption{Relative behavior of the phase $\Theta(r)$ of the SDW
(dashed line) and the shock front (full line) for the four cases of Fig.\,1.}
\end{figure}

\begin{figure}
\caption{Relative behavior of the two arm phase $\Theta_B(r)$ 
and $\Theta_I(r)$
for NGC\,7479. The diagram indicates the
presence of two CRs at $r\approx 23"$ and $r\approx 55"$. The SDW
is trailing around the inner CR and leading around the outer one.}
\end{figure}

\begin{figure}
\caption{CR circles on  NGC\,7479  B image. N at the top and E to the left.
Scales are in pixels, 1\,pxl=0.560'', V$_R$=2378\,km/sec.}
\end{figure}

\begin{figure}
\caption{Relative behavior of the phases $\Theta_B(r)$ and
$\Theta_I(r)$ for NGC\,1832. The diagram indicates the presence of
three CRs at $r\approx 15"$, $r\approx 34"$, and $r\approx 40"$.
The SDW is leading around the inner and
the outer CRs and it is trailing around the intermediate one.}
\end{figure} 

\begin{figure}
\caption{B image of NGC\,1832 showing the position of the three CRs. 
N at the top and E to the left.
Scales are in pixels, 1\,pxl=0.414'', V$_R$=1937\,km/sec.}
\end{figure} 

\begin{figure}
\caption{We show schematically for NGC\,1832,  
the possible relative 
behavior of the  disk (full line) and the perturbing pattern 
(dashed line)  angular velocities.}  
\end{figure}

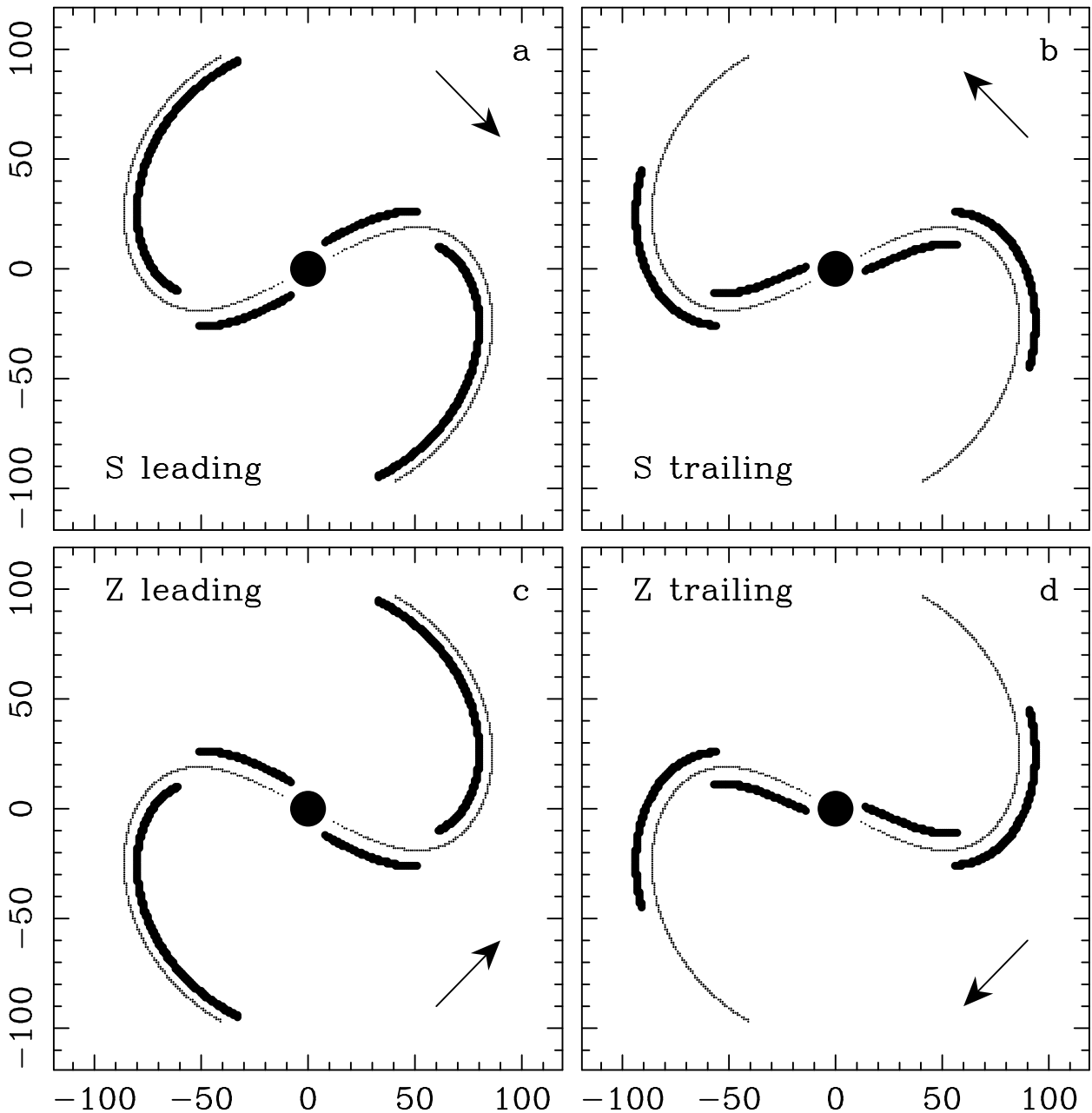
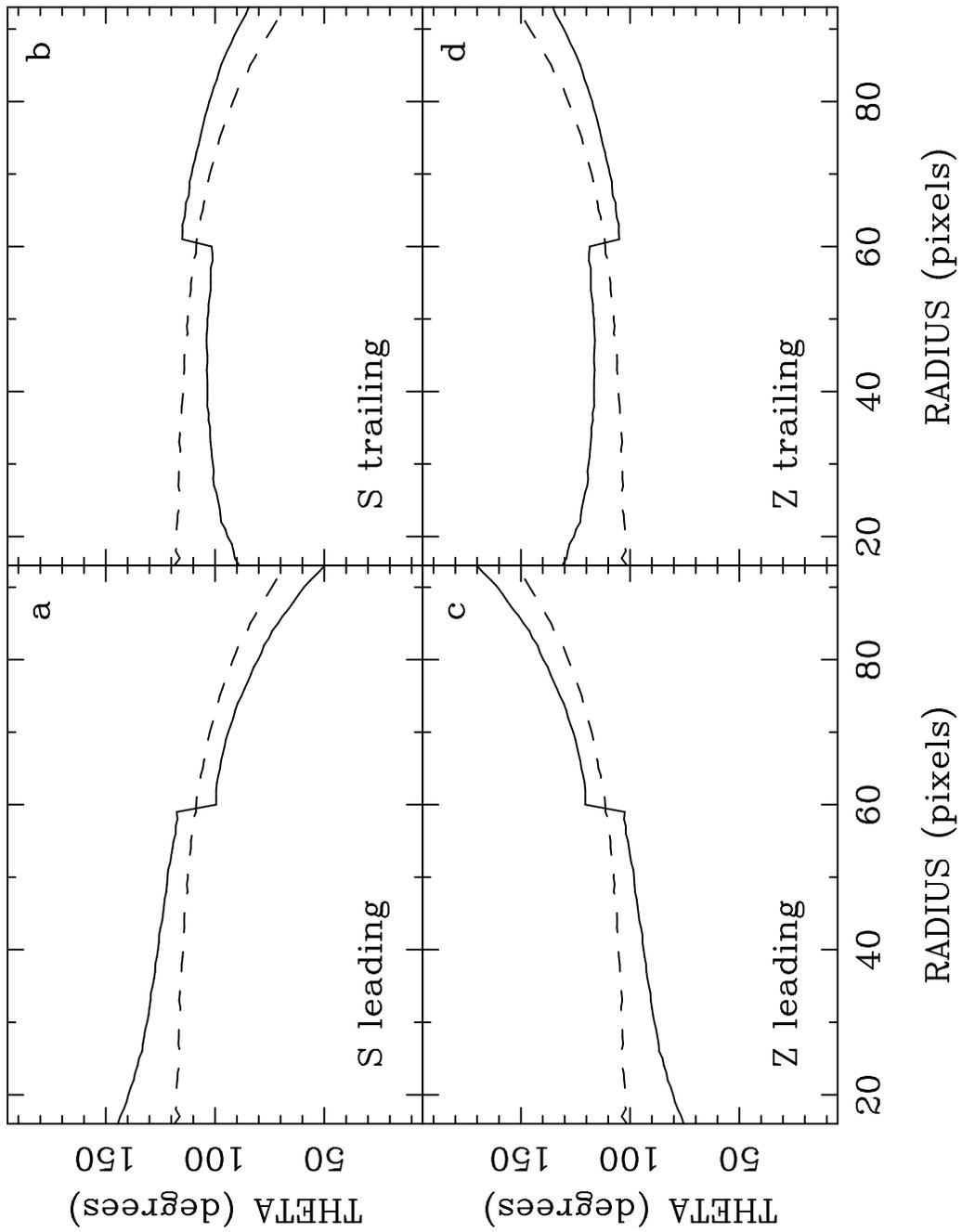
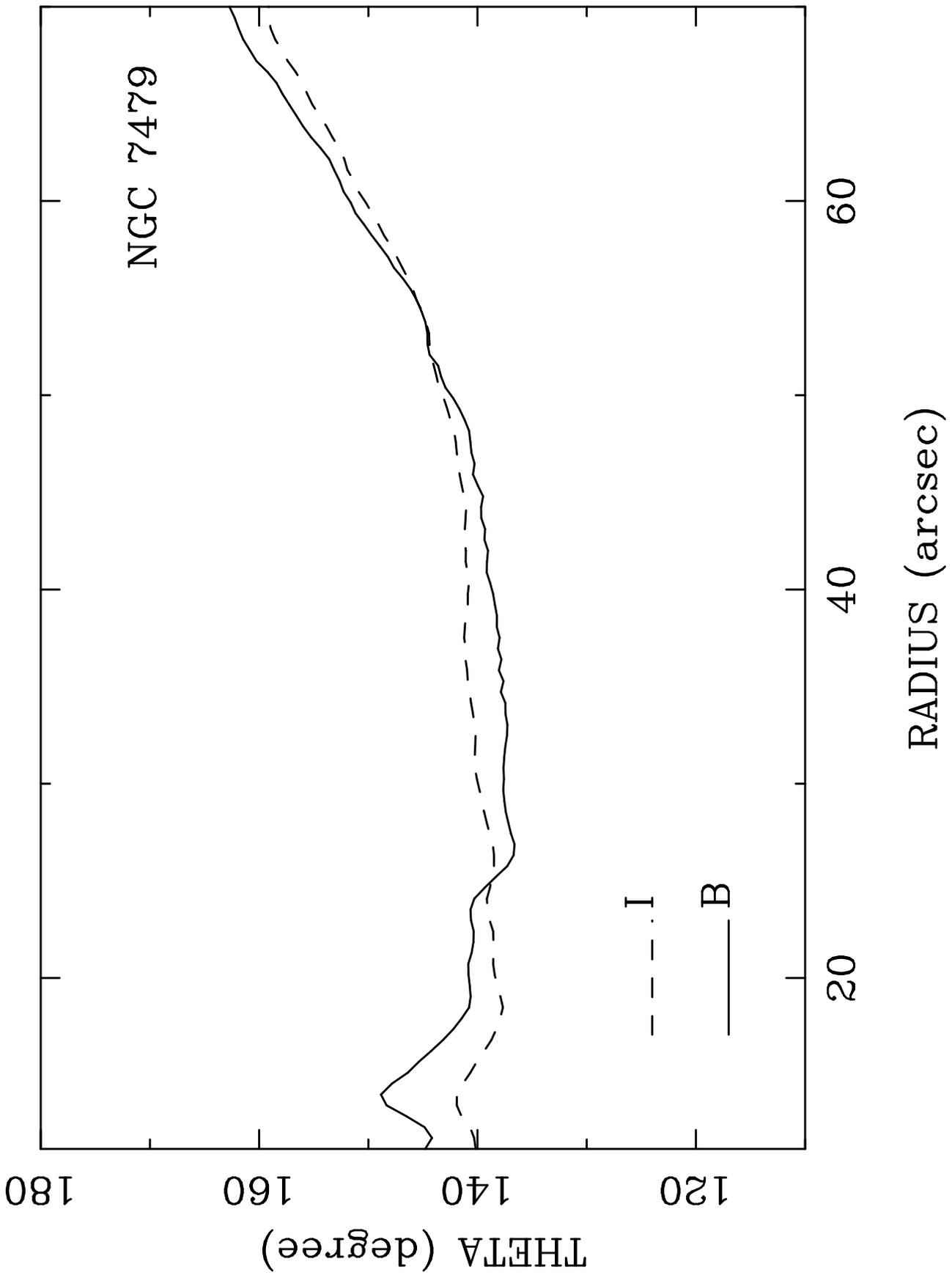
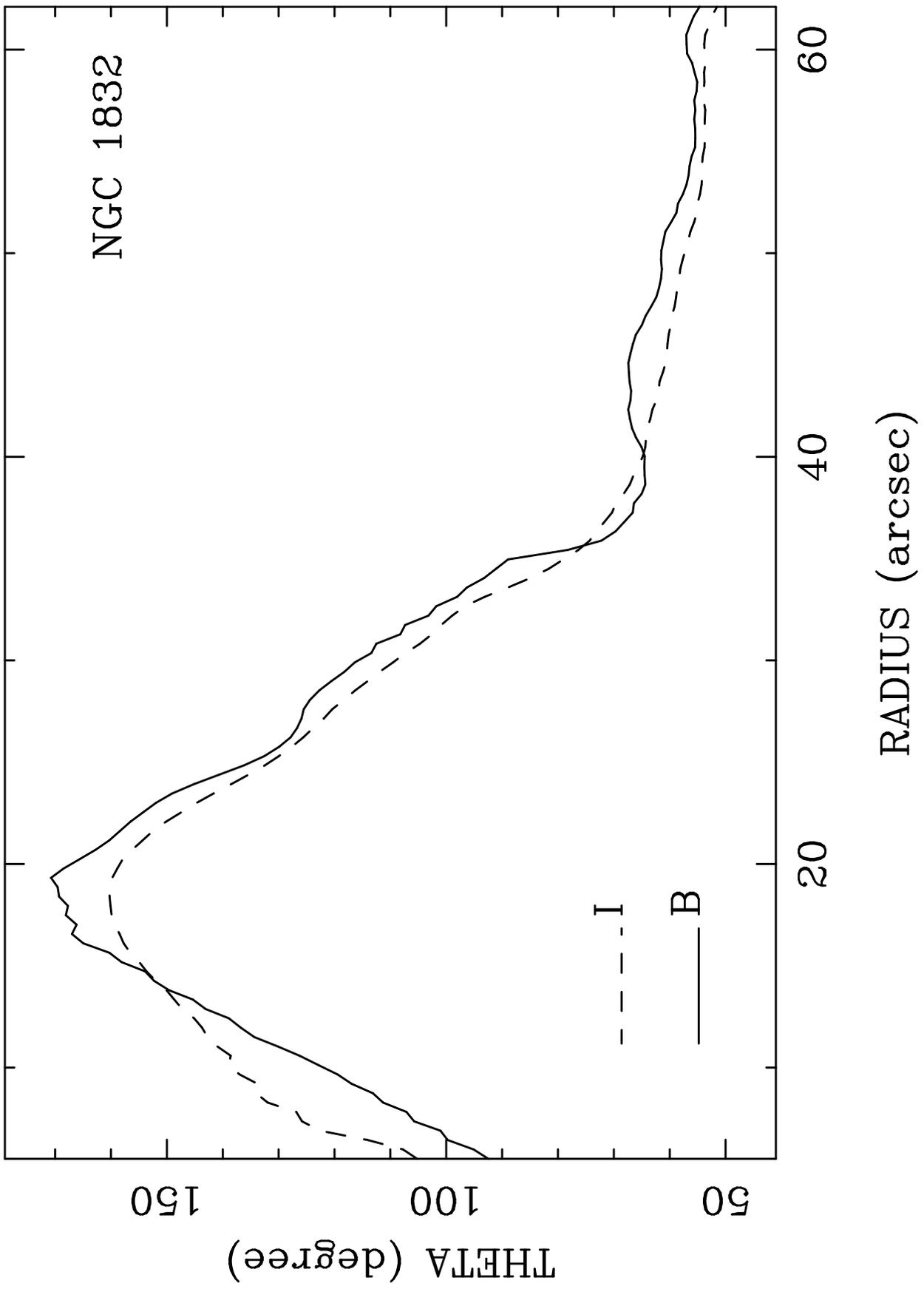
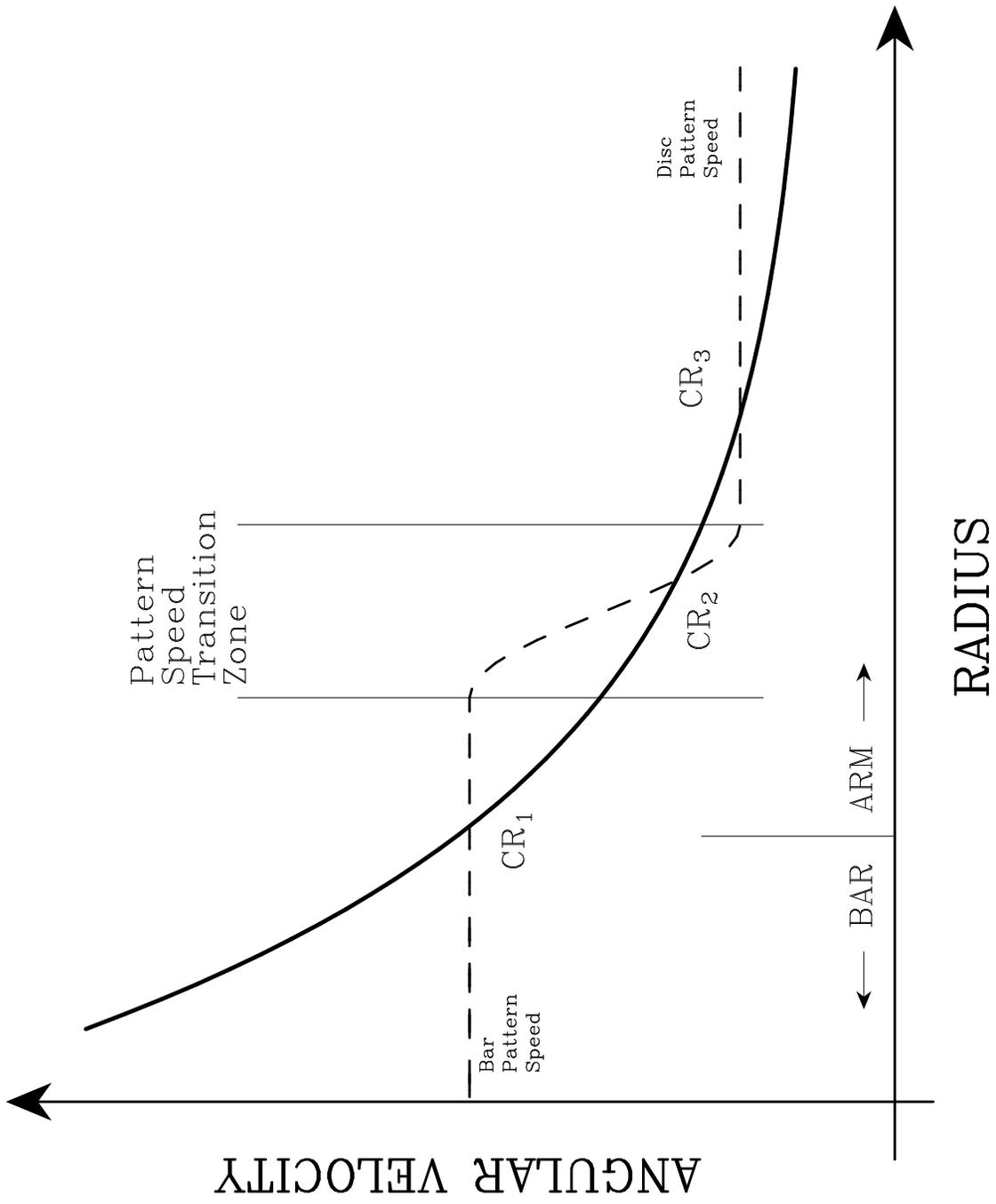
\end{document}